\documentclass[preprint]{elsarticle} 
\pdfoutput=1
\usepackage{amsmath}
\usepackage{amssymb}
\usepackage{subcaption}
\usepackage{multicol}
\usepackage{xcolor}
\newcommand{\bea}{\begin{eqnarray}}
\newcommand{\eea}{\end{eqnarray}}
\newcommand{\nn}{\nonumber}

\begin{document}

\begin{frontmatter}
\title{Exploring vector dark matter via effective interactions}
\author[1]{Hrishabh Bharadwaj \corref{cor1}%
\fnref{fn1}}
\ead{hrishabhphysics@gmail.com}
\cortext[cor1]{Corresponding author}

\affiliation[1]{organization={Government Mahila Degree College (Rajkiya Mahila Mahavidyalaya)},
addressline={Budaun},
postcode={243601, Affiliated to Mahatma Jyotiba Phule Rohilkhand University, Bareilly},
city={Uttar Pradesh},
country={India}} 

\begin{abstract}
We explore the properties of vector dark matter (DM) particles that predominantly interact with Standard Model (SM) electroweak gauge bosons, using an effective field theory approach. The study emphasizes effective contact interactions, invariant under the SM gauge group, between vector DM and SM-neutral electroweak gauge bosons. Focusing on interaction terms up to dimension-8, we establish constraints on the model parameters based on the observed DM relic density and indirect detection signals. We also examine the prospects for dark matter-nucleon scattering in direct detection experiments. In addition, we analyze the sensitivity of low-energy LEP data to the pair production of light DM particles (with masses up to 80 GeV). Finally, we assess the potential of the proposed International Linear Collider (ILC) to probe these effective operators through the detection of DM particles produced in association with mono-photons.
\end{abstract}

\begin{keyword}
Vector dark matter \sep Collider \sep ILC
\end{keyword}

\end{frontmatter}



\section{Introduction}
Dark matter (DM) accounts for approximately 26.5\% of the Universe's energy density and 84.1\% of its total matter, as indicated by various cosmological and astrophysical observations. Precise measurements of DM density from the Planck Collaboration estimate the relic density to be $\Omega_{DM} h^2 = 0.1198 \pm 0.0012$ \cite{Planck:2018vyg}. Despite significant advancements in our understanding, the true nature of DM remains unknown.

Efforts to detect DM are underway in multiple areas. Direct detection experiments, such as  DarkSide-50 \cite{DarkSide:2018kuk}, DarkSide-20k \cite{Manthos:2023swh}, LUX-ZEPLIN (LZ) \cite{LZ:2019sgr}, PandaX-4T \cite{PandaX-4T:2021bab},  XENONnT \cite{XENON:2020kmp}, CRESST \cite{CRESST:2016qpj}, LUX \cite{LUX:2016ggv},DAMA/LIBRA \cite{Bernabei:2013xsa, Bernabei:2018yyw}, , CoGeNT \cite{CoGeNT:2012sne},PandaX-II \cite{PandaX-II:2017hlx}, and CDMS \cite{CDMS:2013juh} aim to observe the recoil of atoms or nucleons as a result of DM interactions. Collider experiments, both current and future, concentrate on detecting DM through monojet or dijet events associated with missing energy. Meanwhile, indirect detection experiments like  AMS-02 \cite{AMS:2014xys, AMS:2016oqu}, FermiLAT \cite{Fermi-LAT:2015att}, and HESS \cite{HESS:2013rld} search for excess in cosmic rays flux coming from DM annihilation into SM particles.

The Effective Field Theory (EFT) approach offers a model-independent framework for studying DM phenomenology by treating SM-DM interactions as contact interactions expressed by non renormalizable operators \cite{Chen:2013gya, Chae:2012bq,Rajaraman:2012fu,Farzan:2012hh,Arcadi:2020jqf,Krnjaic:2023nxe,Nomura:2024jea,Bharadwaj:2020aal,Singh:2024wdn,Tran:2023lzv,Bertuzzo:2024bwy,Bhattacharya:2022qck}. This formalism enables exploration of DM phenomenology across various processes, including annihilation and scattering. Several studies have constrained the parameter space for quark-DM and gauge boson-DM interactions at the Large Hadron Collider (LHC), using simplified models and other popular scenarios \cite{Cotta:2012nj, Crivellin:2015wva, Chen:2015tia, Bell:2012rg,Bharadwaj:2021tgp}.In a recent paper \cite{Aebischer:2022wnl}, the authors extend the SMEFT and LEFT by incorporating additional SM singlet vector DM particles of mass up to $10$ GeV. They classify all gauge-invariant interactions up to dimension-six terms and provide tree-level matching conditions between SMEFT and LEFT at the electroweak scale. They demonstrate the model's viability through freeze-in production mechanism.

In the context of deep inelastic scattering of leptons and hadrons, analyses of twist-2 tensor operators and their renormalization group equations have provided insights into DM-nucleon scattering, with contributions from hadronic operators. Further research into one-loop effects from twist-2 quark and gluonic operators has been conducted, enhancing our understanding of DM-nucleon interactions \cite{Drees:1993bu, Hisano:2010ct, Hisano:2010yh, Hisano:2017jmz}. Collectively, these efforts contribute to ongoing investigations into DM through direct and indirect detection, as well as collider searches.

In this paper, we study vector DM particles interacting primarily with the neutral SM electroweak gauge bosons, \emph{viz.} $Z\ \&\ \gamma$, via effective operators within the $SU(2)_L \times U(1)_Y$ gauge framework. We consider a real vector DM candidate, focusing on three types of operators, all of which are SM gauge singlets.

In Section \ref{model}, we construct the effective Lagrangian for the interactions of vector DM and SM neutral electroweak gauge bosons, considering operators of higher dimensions. Section \ref{pheno} presents constraints on the model parameters based on relic density measurements and consistency checks from direct and indirect detection experiments. Section \ref{coll} discusses the constraints derived from Large Electron-Positron Collider (LEP) data and assesses the sensitivity of these operators at the proposed International Linear Collider (ILC). Finally, we summarize our findings in Section \ref{summ}.
\section{Effective Lagrangian}
\label{model}
\par The SM is extended with a set of effective operators up to dimension 8. The effective Lagrangian describing the interaction between a real vector DM particle ($V$) and the SM neutral electroweak gauge bosons is given as follows: 

\bea
{\mathcal L}_V&=&\frac{{\alpha^{V}_{S_1}}}{\Lambda^4}\ { O}_{S_1}^{V} + \frac{{\alpha^{V}_{S_2}}}{\Lambda^2}\ {O}_{S_2}^{V} + \frac{{\alpha^{V}_{T}}}{\Lambda^4}\ { O}_{T}^{V}, \label{LVDM}
\eea
where $\Lambda$ represents cut-off scale of the effective theory  and $\alpha_i^V\ (i =\ S_1,\ S_2,\ T)$ is the respective strength of the effective interaction. The effective operators considered for vector DM are given by:
\bea
 {O}^{V}_{S_1}&\equiv& m_{V}^2\  {V}^\mu \, {V}_\mu\ B_{\mu\nu}\ B^{\mu\nu} \nn\\
{O}^{V}_{S_2}&\equiv&  {V}^\rho \, {V}_\sigma\ B_{\mu\rho}\ B^{\mu\sigma} \nn\\
{O}^{V}_{T}&\equiv& \ {V}^\rho\,\, i\, \partial^\mu \,\,i\,\partial^\nu\,\, {V}_\rho\,\, {O}_{\mu\nu} \ +\ \text{h.c.},
\label{Operators}
\end{eqnarray}

where $O_{\mu\nu}\ =\ B^\alpha_\mu\ B_{\alpha \nu}\ -\ \frac{1}{4}\ g_{\mu \nu}\ B^{\alpha \beta}\ B_{\alpha \beta}$, and the field strength tensor for the SM $U(1)_Y$ gauge is denoted by $B_{\mu\nu}$.

\section{Dark matter phenomenology}
\label{pheno}
\subsection{Constraints from relic density}
\par In the early Universe, DM particles remained in thermal equilibrium with the plasma through continuous annihilation and production processes. As the Universe expanded and the DM particles became non-relativistic, they gradually deviated from thermal equilibrium. Finally, they \textit{froze out} and persisted as a cold relic when their annihilation rate dropped below the Hubble expansion rate. The Boltzmann equation describes this evolution, and the approximate expression for the DM relic density is as follows \cite{Bauer:2017qwy,Dodelson:2003ft}:

\bea
\Omega_{V}\ h^2 &\approx& 0.12\ \sqrt{\frac{g_\star(x_f)}{100}}\ \left(\frac{x_f}{28}\right)\ \frac{2\times 10^{-26}\ cm^3/s}{\langle\sigma\	v\rangle},
\eea
where $\langle \sigma\ v\rangle,\ g_\star(x_f),\ \&\ x_f $ represent the thermally averaged DM annihilation cross-section, effective degrees of freedom at freeze-out, and the ratio of DM mass and temperature at freeze-out, respectively. The thermally averaged cross-section can be expressed in terms of relative velocity of the DM particles: $\langle \sigma\ v \rangle\ =\ a\ +\ b\ \langle v^2 \rangle\ +\ c\ \langle v^4 \rangle +\mathcal{O}(v^6)$, where $\langle v^2 \rangle\ =\ \frac{6}{x_f}\ \&\ \langle v^4 \rangle\ =\ \frac{60}{x_f^2}$. The expressions for thermally averaged cross-sections corresponding to the processes contributing to the DM relic density are provided in Appendix \ref{App}.

\begin{figure}[ht]
\centering
  \includegraphics[width=\linewidth]{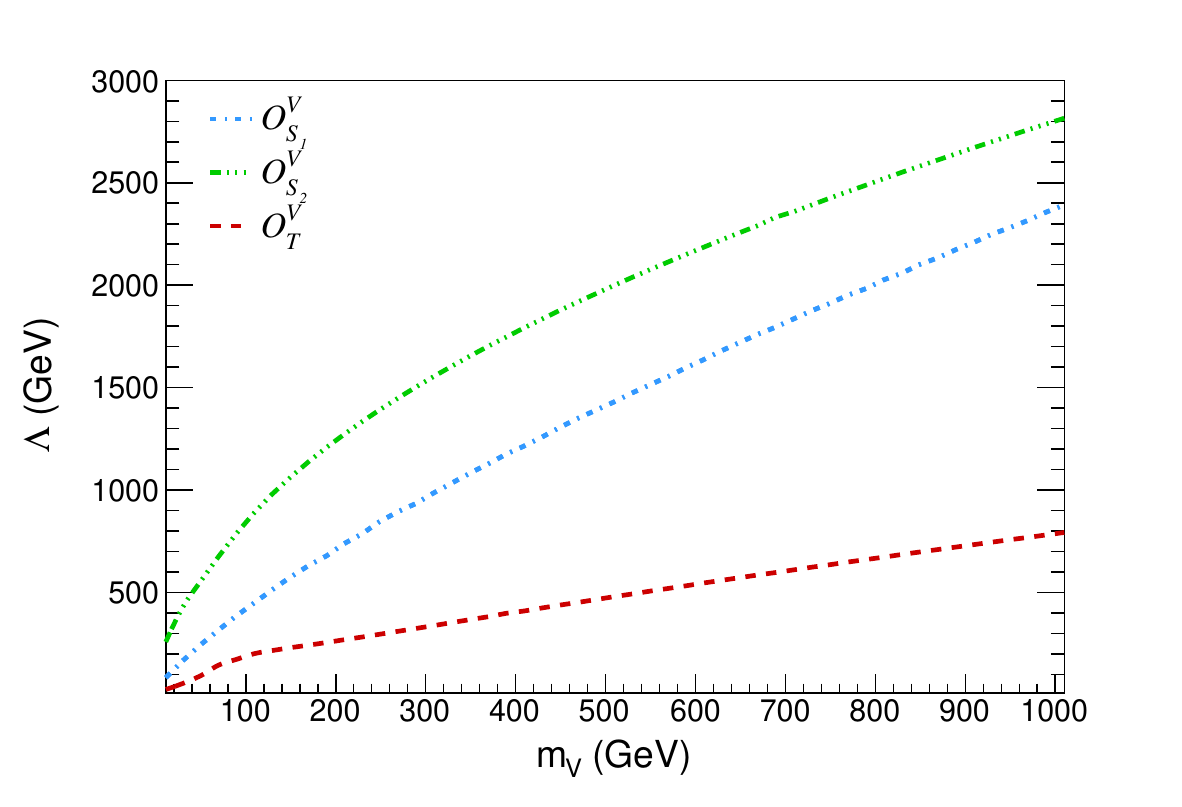}
\caption{The contour lines correspond to DM relic density $\Omega\ h^2 = 0.1198$ for the associated DM operators. The parameter space under the respective contour lines is allowed from the observed relic density.}
\label{fig1}
\end{figure}

\par We utilized FeynRules \cite{Alloul:2013bka} to create the couplings, vertices and additional model files that are required. The relic density was then calculated using these files in the MadDM \cite{MadDM} package. Figure \ref{fig1} shows contour plots in the $\Lambda - m_{V}$ plane, where the current DM relic density of $\Omega\ h^2\ =\ 0.1198$ \cite{Planck:2018vyg} is satisfied for DM masses between 10 GeV and 1 TeV. The areas under the respective contour curves are consistent with constraints from the relic density. In all relic density calculations, the couplings $\alpha_i$ were set to 1. Additionally, we have turned on each operator separately.

\subsection{Constraints from indirect detection}
The annihilation of DM in dense regions of the galaxies would produce a significant amount of high energy SM particles. The High Energy Stereoscopic System (HESS) \cite{HESS:2018zix} employs indirect detection methods to search for DM by looking for high-energy gamma rays coming from particular areas, including dwarf galaxies or the galactic core. HESS aims to identify possible signals resulting from DM annihilations or decays, which can generate gamma rays as secondary particles when DM interacts and converts into SM particles. By accurately measuring gamma-ray fluxes and energy spectra, HESS provides interesting observations of the properties and distribution of DM particles, helping to constrain the parameter-space.

The anticipated annihilation rate at the Cherenkov Telescope Array (CTA)\cite{Hofmann:2023fsn} depends on the DM profiles but scales with the cross section. Although they differ by annihilation channel, gamma-ray spectra are accurately predicted for each. We calculate the DM annihilation cross sections into a pair of photons, which are provided below:

\begin{eqnarray}
\sigma^{V}_{S_1}\ v\ \left(V\ V \to \gamma \gamma\right)&\simeq&
\frac{16\ {\alpha^{V}_{S_1}}^2\ }{9 \pi}\ \cos^4\theta_w \ \frac{m_{V}^6}{\Lambda^8}\ \left(7\ +\ \frac{5}{3}\ v^2\right)
\label{indS1}\\
 \sigma^{V}_{S_2}\ v\ \left(V\ V \to \gamma \gamma\right)&\simeq&
\frac{{\alpha^{V}_{S_2}}^2\ }{9 \pi}\ \cos^4\theta_w \ \frac{m_{V}^2}{\Lambda^4}\ \left(7\ +\ \frac{5}{3}\ v^2\right)
\label{indS2}\\
\sigma^{V}_{T}\ v\ \left(V\ V \to \gamma \gamma\right)&\simeq&
\frac{16\ {\alpha^{V}_{T}}^2\ }{27 \pi}\ \cos^4\theta_w \ \frac{m_{V}^6}{\Lambda^8}\ v^2\label{indT}
\end{eqnarray}
\begin{figure}[ht]
\centering
  \includegraphics[scale=0.7]{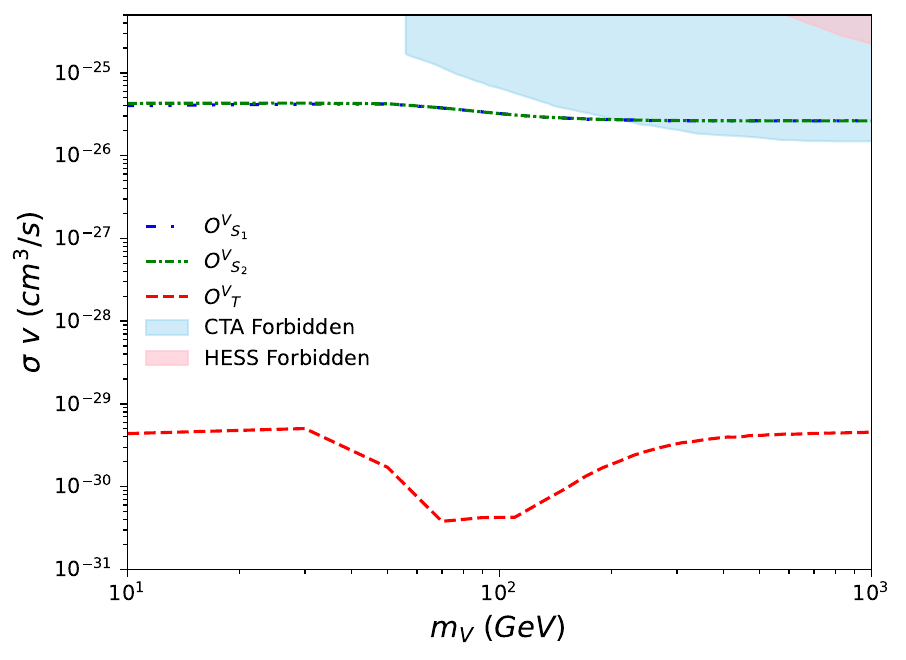}
\caption{DM annihilation into $\gamma\ \gamma$: cross-section versus DM mass, with cut off $\Lambda$ allowed from the observed relic density corresponding to each operator. Based on HESS \cite{HESS:2018zix} and CTA \cite{Hofmann:2023fsn} data, the shaded area is not allowed respectively.}
\label{fig3}
\end{figure}
We scan the DM mass between 10 GeV to 1 TeV.

We scan the DM mass between 10 GeV to 1 TeV. In the indirect detection cross sections provided in equations \eqref{indS1}-\eqref{indT}, the terms independent of the DM relative velocity $v$ arise purely from s-wave annihilation, while the terms proportional to $v^2$ correspond to p-wave annihilation. We assume a DM relative velocity of approximately $10^{-3}\ c$. The equations in \eqref{indS1}-\eqref{indT} indicate that $O^V_{S_1}$ and $O^V_{S_2}$ operators exhibit s-wave behavior, while the operator $O^V_T$ exhibits $p-$wave suppression. When calculating the DM annihilation cross-section into photon pairs, we use parameters consistent with relic density constraints for the given DM mass $m_V$. For each $m_V$, the corresponding annihilation cross section $\sigma v$ is shown in Fig. \ref{fig3}. The region above each line is allowed, while the shaded area represents the exclusion zone from HESS and CTA indirect detection limits. A dip appears in Fig. \ref{fig3} around \( m_{V} \approx 45 - 90 \) GeV due to the opening of the \( Z\gamma \) and \( ZZ \) annihilation channels. The resulting increase in the total annihilation cross-section reduces the DM number density. Consequently, for a fixed DM relic density and coupling constant, the contour in the \( m_{\text{DM}} - \Lambda \) plane (Fig. \ref{fig1}) exhibits a bump, which ultimately manifests as a dip in the \( m_V - \sigma v \) plane.

\subsection{DM-nucleon interaction}

\begin{figure}[htb]
\centerline{\includegraphics[scale=0.8]{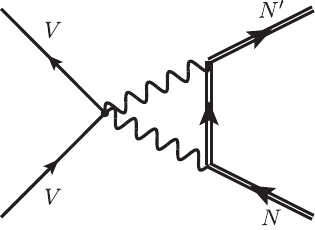}}
\caption{Feynman diagram depicting the scattering of dark matter particles with nucleons.}
\label{DDloop}
\end{figure}

The goal of direct detection experiments is to investigate how DM particles scatter off of nucleons or atoms \cite{DarkSide:2018kuk,Manthos:2023swh,LZ:2019sgr,PandaX-4T:2021bab,LUX:2016ggv,Bernabei:2013xsa, Bernabei:2018yyw,CoGeNT:2012sne,PandaX-II:2017hlx,CDMS:2013juh}. The purpose of these probes is to measure the recoil momentum of atoms or nucleons inside the detector material. DM-nucleon, DM-atom, and DM-electron interactions are the three general types into which scattering events occur. Such interactions can only take place at loop levels in our model, as DM does not directly interact with leptons, quarks, or gluons at the tree level.Figure \ref{DDloop} shows this for DM-nucleon scattering, which is the next focus of our current study.

\section{Collider searches}
\label{coll}
\subsection{Constraints from LEP}
By utilizing existing findings and observations from LEP data, we can establish constraints on effective operators. The cross-section for the process \(e^+e^-\to \gamma^\star + V V\) is compared with the combined analysis conducted by the DELPHI and L3 collaborations for \(e^+e^- \to \gamma^{\star} + Z \to q \bar{q} + \nu_{l}\bar{\nu}_{l}\) at a center-of-mass energy \(\sqrt{s} = 196.9\) GeV, with an integrated luminosity of 0.679 fb\(^{-1}\). Here, \(q\) represents the light quarks \(u\), \(d\), and \(s\), while \(\nu_{l}\) denotes the SM neutrinos, including \(\nu_e\), \(\nu_{\mu}\), and \(\nu_{\tau}\).

\begin{figure}[h]
\centerline{\includegraphics[scale=0.6]{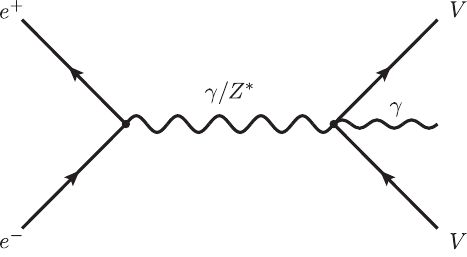}}
\caption{Feynman diagram illustrating the production of a DM pair associated with a photon in an $e^+ e^-$ collider.}
\label{FeynD}
\end{figure}

\begin{figure}[ht]
\centering
  \includegraphics[width=\linewidth]{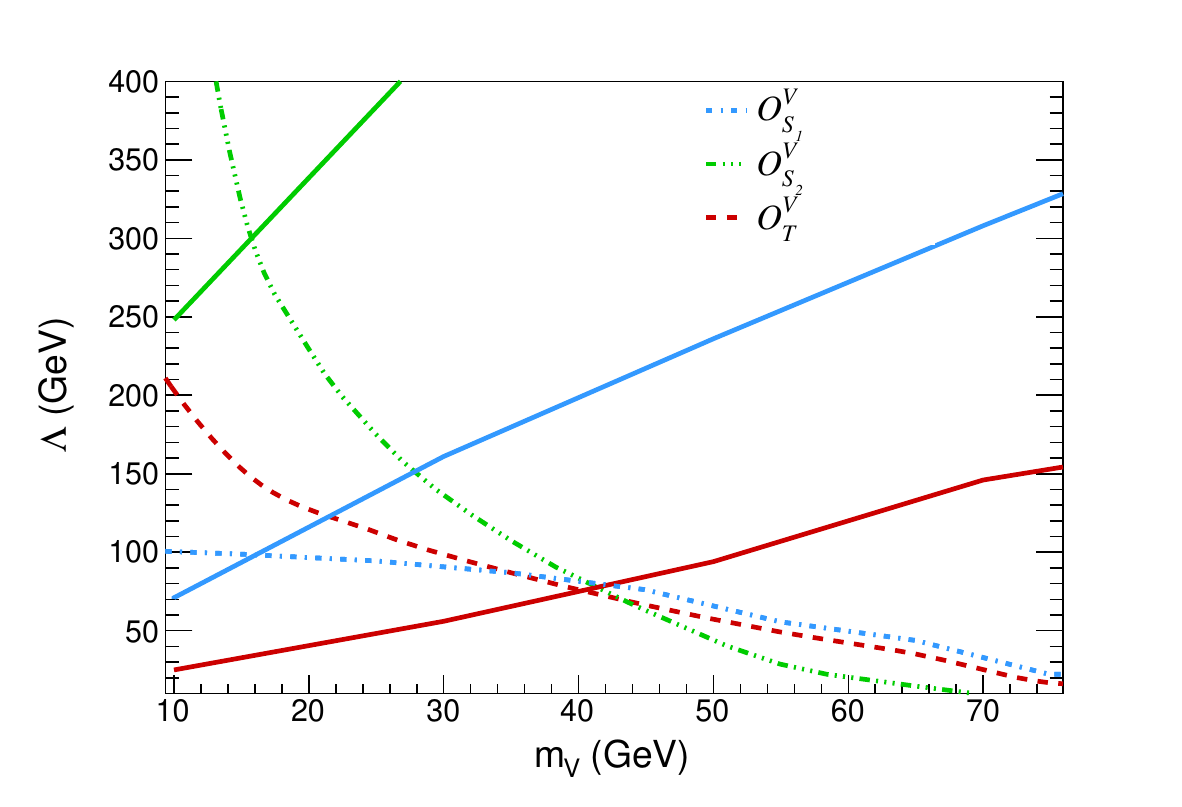}
\caption{Dashed lines represent the contours within the parameter space defined by the DM mass and the kinematic reach for $e^+e^- \to VV + \gamma^\star \to \,\,\not\!\!\!E_T + q_{i}\bar q_{i}$ at a center of mass energy of $\sqrt{s} = 196.9$ GeV, with an integrated luminosity of 679.4 pb$^{-1}$. These contours correspond to the constraint $\delta\sigma_{\rm tot} = 0.032$ pb, obtained from a combined analysis by DELPHI and L3 \cite{ALEPH:2013dgf}. The region below the dashed lines is not allowed based on LEP observations. Moreover, the areas below the solid lines, corresponding to specific operators, satisfy the relic density requirement $\Omega_{\rm DM} h^2 \leq 0.1198$.}
\label{fig5}
\end{figure}

The Feynman diagram for the production of a photon \(\gamma\) with missing energy, arising from effective operators as described in equation \eqref{Operators}, at the {\color{red}\(e^- e^+\)} lepton collider is shown in Figure \ref{FeynD}. The measured cross section for this process is \(0.055\) pb, with a statistical uncertainty (\(\Delta\sigma_{\text{stat.}}\)) of \(0.031\) pb, a systematic uncertainty (\(\Delta\sigma_{\text{sys.}}\)) of \(0.008\) pb, and a total uncertainty (\(\Delta\sigma_{\text{total}}\)) of \(0.032\) pb, as reported in \cite{ALEPH:2013dgf}. Therefore, the contribution from any new physics involving DM pair, resulting in missing energy and two quark jets, can be accommodated in the observed \(\Delta\sigma_{\text{total}}\). In Figure \ref{fig5}, we present dashed line contours at a 95\% confidence level that correspond to \(\Delta\sigma_{\text{total}} \simeq 0.032\) pb, related to the operators in the DM mass-\(\Lambda\) plane. The region below the dashed lines, as indicated, is excluded by the combined analysis of LEP, while the solid lines represent the contours from relic density.

\subsection{Analysis at the ILC}

In this subsection, we examine DM pair production plus mono photon processes at the future collider, International Linear Collider (ILC) for the DM mass between 10 and 240 GeV. The specific process we consider is \(e^+\,e^-\,\rightarrow\, V\,V\,\gamma\), as illustrated in Figure \ref{FeynD}. The main SM background for the \(e^+e^-\to \not \!\! E_T\,\gamma\) process originates from \(e^+\,e^-\rightarrow\, Z\,\gamma \to \nu\,\overline{\nu}\,\gamma\).

\begin{table*}[htb]\footnotesize
\centering
\begin{tabular*}{\textwidth}{c|@{\extracolsep{\fill}} cccc|}\hline\hline
	&\textit{ILC-250}&\textit{ILC-500}\\
	$\sqrt{s} \left( \textit{in GeV}\right )$& 250 & 500   \\
	$L_{int} \left( \textit{in $fb^{-1}$}\right )$ & 250 & 500   \\
	$\sigma_{BG} \,(pb)$ &1.07 &  1.48 \\\hline\hline
\end{tabular*}
\caption{\small \em{Accelerator parameters for the analysis at ILC.}}
\label{table:accelparam}
\end{table*}

The investigation of signal and background processes, utilizing the accelerator parameters given in the technical design report for the ILC \cite{ Behnke:2013xla,Behnke:2013lya} as presented in Table \ref{table:accelparam}, involves the simulation of DM signal SM background events. Model files are generated in FeynRules \cite{Alloul:2013bka}, and events for the signal and background are produced using MadGraph package \cite{Alwall:2014hca}. For event selection and analysis purpose, we have used MadAnalysis 5 \cite{Conte:2012fm}.

\par Along with the basic selection criteria, which involve cuts on the transverse momentum of the photon ($p_{T_{\gamma}} \geq$ 10 GeV) and the pseudo-rapidity of the photon ($\left\vert \eta_\gamma \right\vert \leq$ 2.5), a selection requirement related to the photon energy for on-shell production is also implemented.

In addition to the basic selection criteria, which include cuts on the transverse momentum of the photon (\(p_{T_{\gamma}} \geq\) 10 GeV) and the pseudo-rapidity of the photon (\(\left\vert\eta_\gamma\right\vert\leq\) 2.5), a selection requirement correspoinding to the photon energy against on-shell $Z$ production is applied. The following relation must be satisfied for such events to be excluded:

\bea
\frac{2\,E_\gamma}{\sqrt{s}} = 1\ -\ \frac{(m_Z^2\ \pm\ 10\ m_Z \Gamma_Z)}{s}
\eea

This means that events are rejected when \(2\,E_\gamma/\sqrt{s}\) falls within the intervals \(\left[0.8,0.9\right]\) and \(\left[0.95,0.98\right]\) for center of mass energies, \(\sqrt{s}=250\ \&\ 500\) GeV, respectively.

\begin{figure*}
\centering
\begin{multicols}{2}
\includegraphics[width=0.49\textwidth,clip]{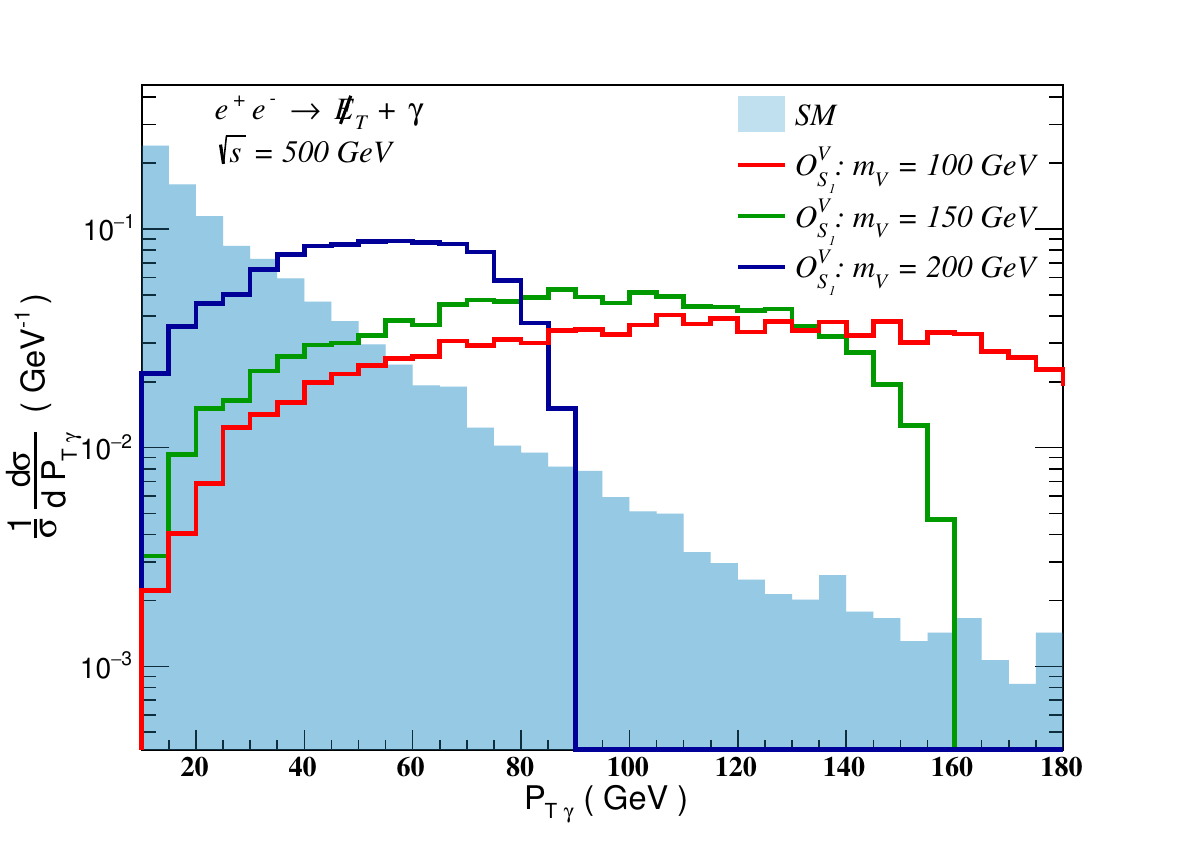}
\subcaption{\small \em{}}\label{ChiSqrFDM1000}
\columnbreak
\includegraphics[width=0.49\textwidth,clip]{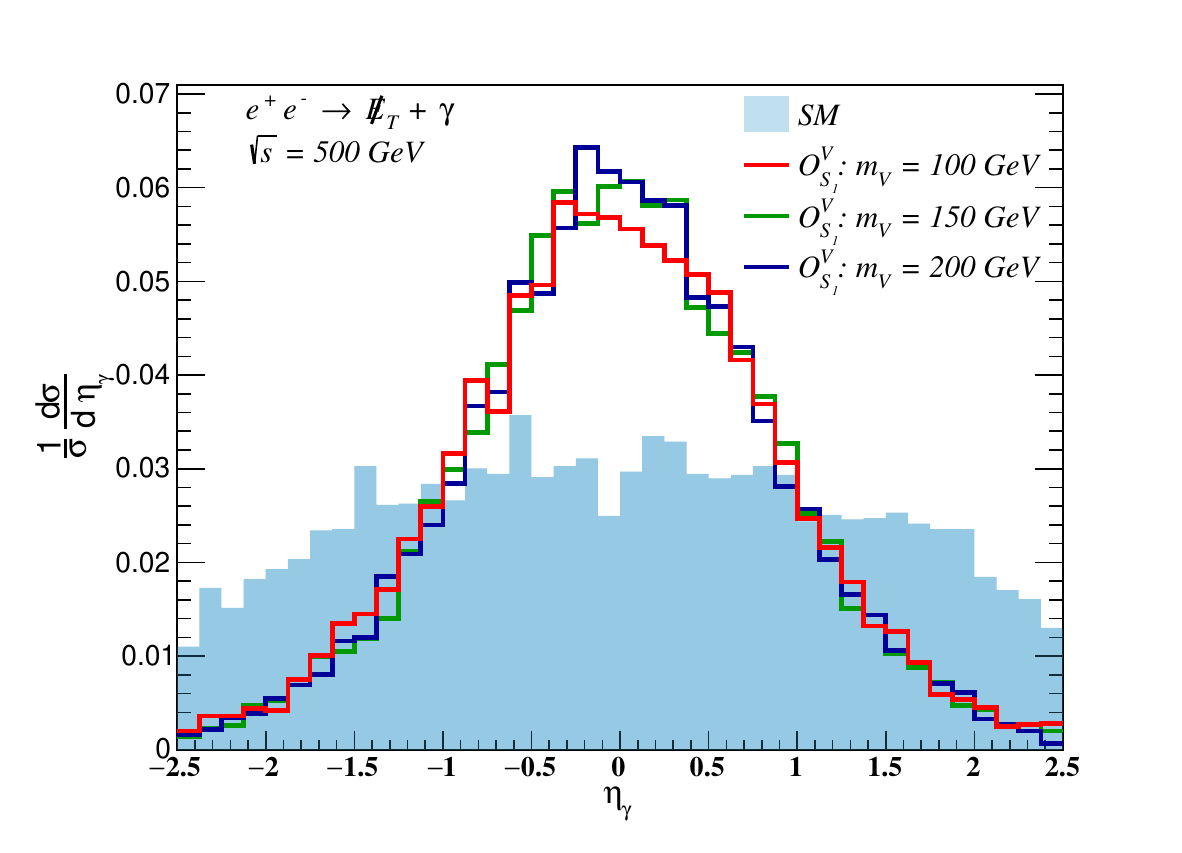}
\subcaption{\small \em{ }}\label{ChiSqrSDM1000}
\end{multicols}
	\begin{multicols}{2}
\includegraphics[width=0.49\textwidth,clip]{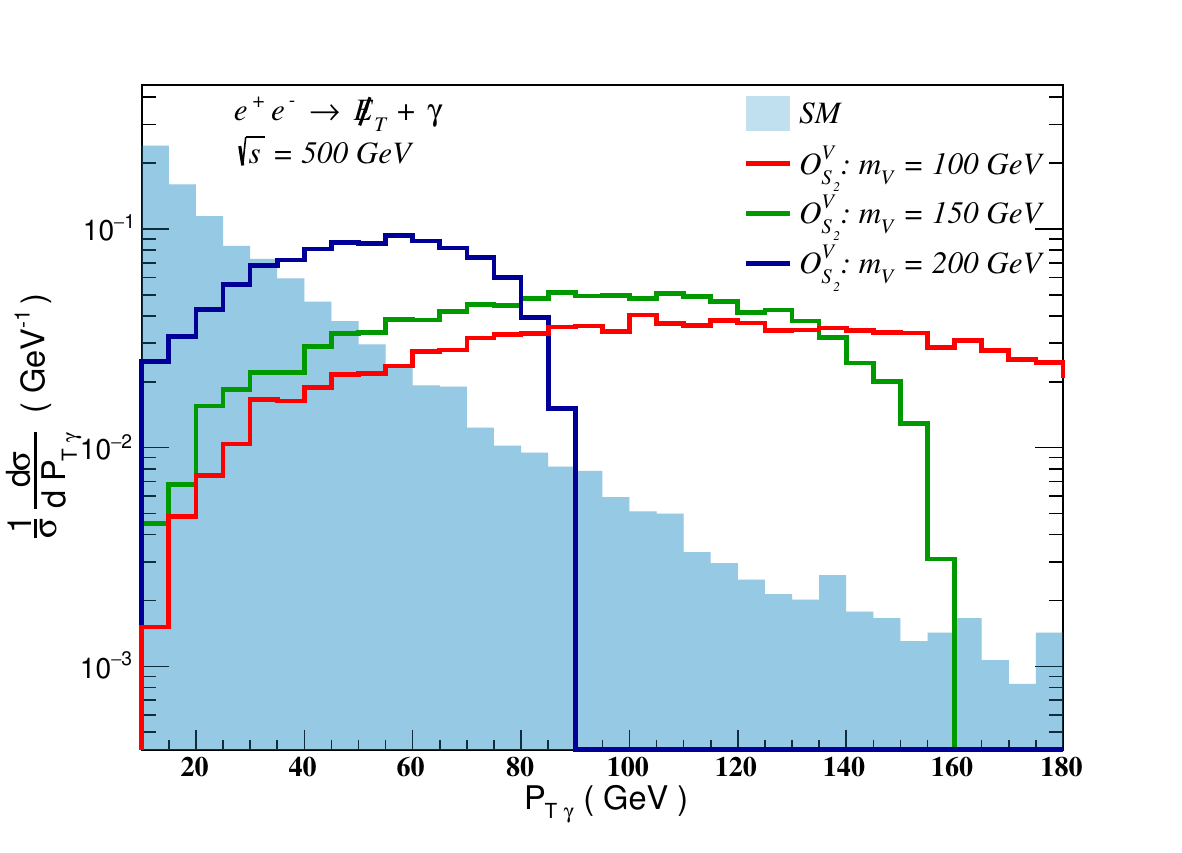}
\subcaption{\small \em{ }}
\columnbreak
\includegraphics[width=0.49\textwidth,clip]{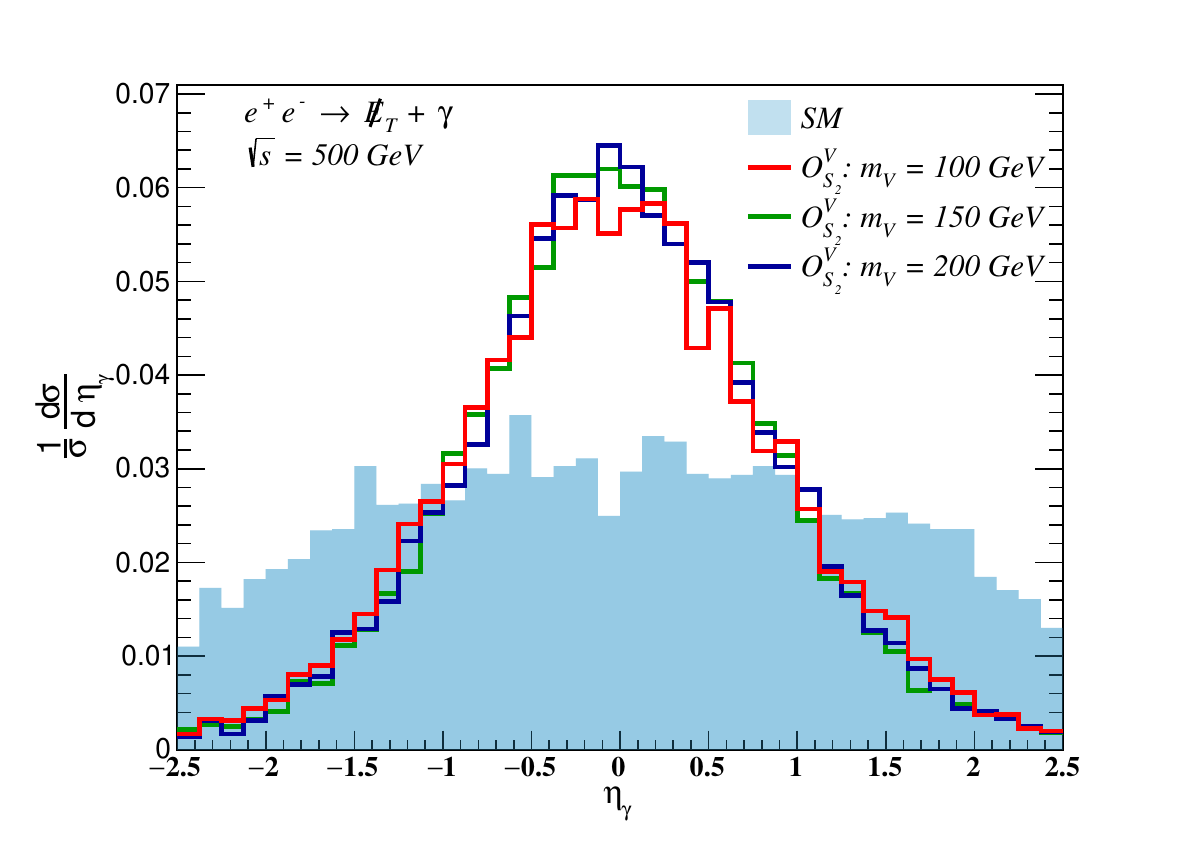}
\subcaption{\small \em{ }}\label{ChiSqrSDM1000}
\label{ChiSqrVDM1000}
\end{multicols}
\begin{multicols}{2}
\includegraphics[width=0.49\textwidth,clip]{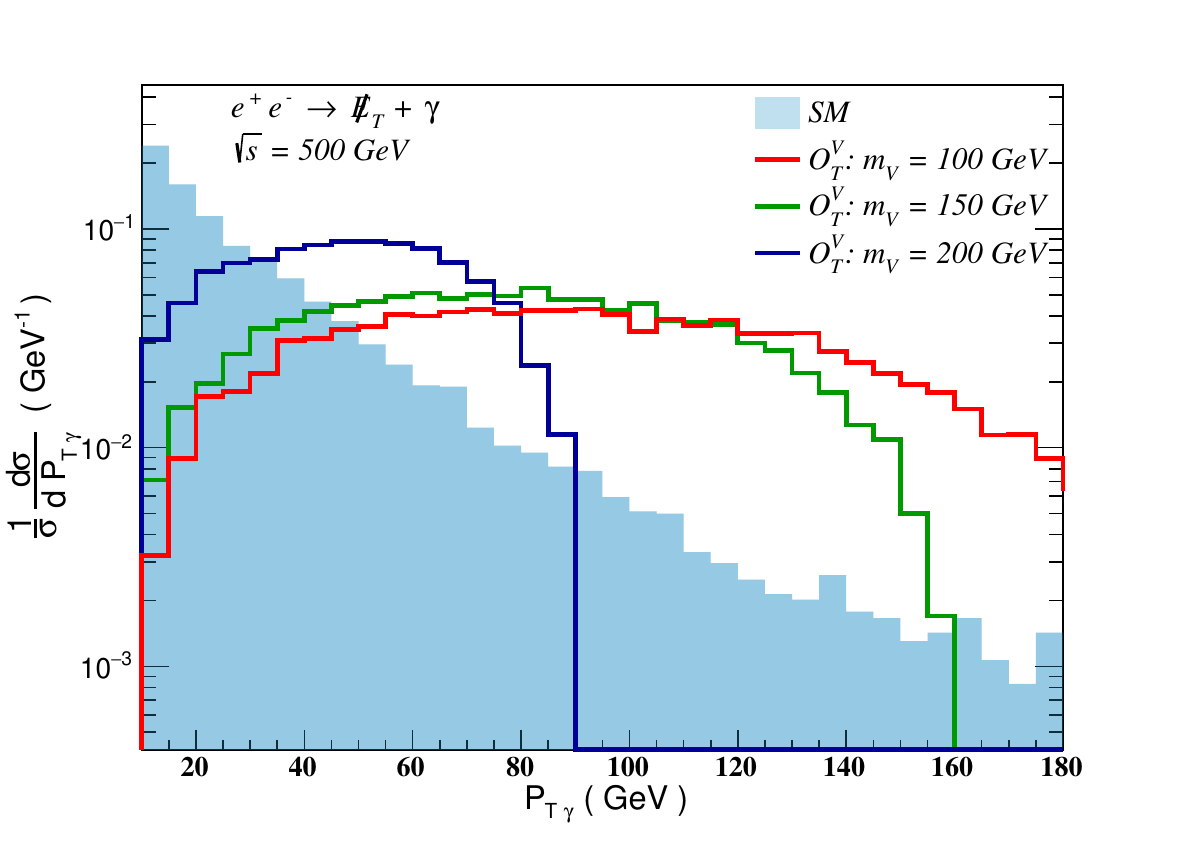}
\subcaption{\small \em{}}
\columnbreak
\includegraphics[width=0.49\textwidth,clip]{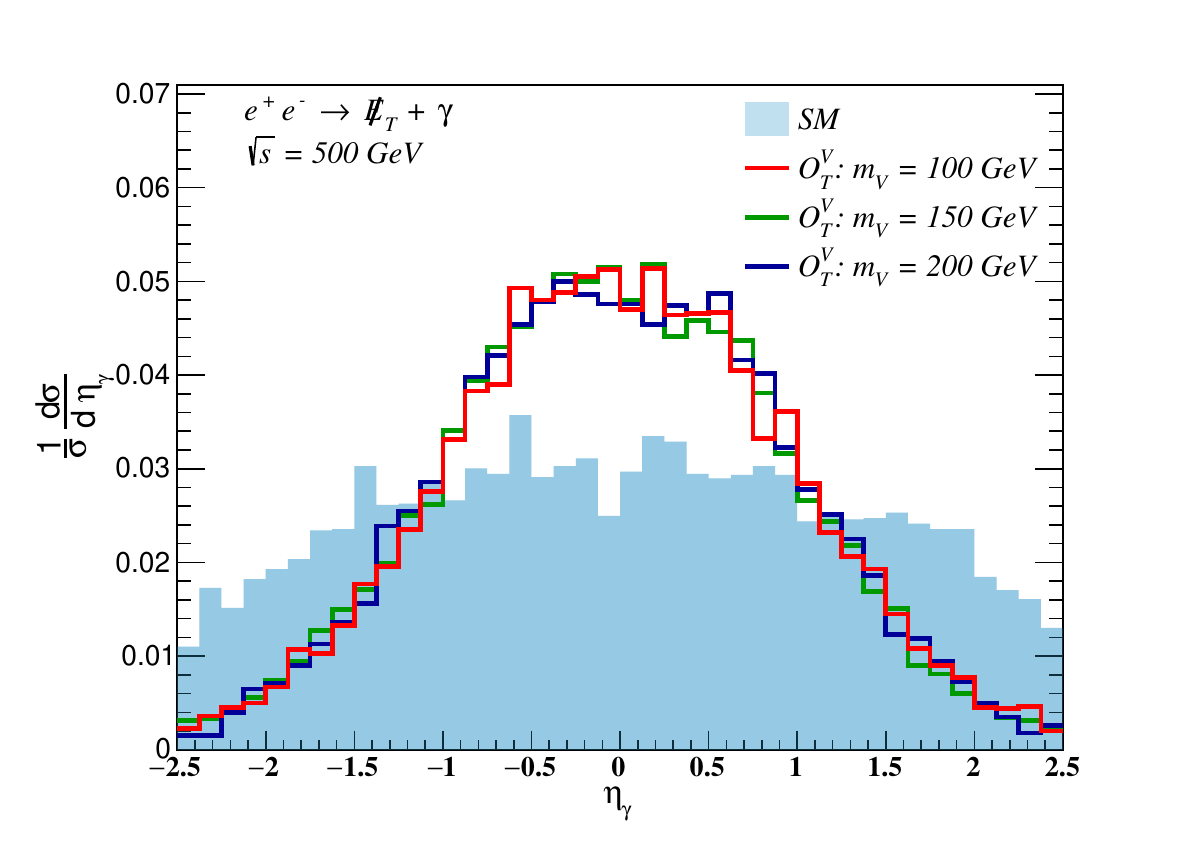}
\subcaption{\small \em{ }}\label{ChiSqrSDM1000}
\label{ChiSqrVDM1000}
\end{multicols}
	\caption{\small \em{We depict the  one dimensional normalized differential cross sections for $p_{T_\gamma}$  and $\eta_{\gamma}$  for both SM processes and those induced by the $O^V_{S_1}$ (upper panel), $O^V_{S_2}$ (middle panel), and $O^V_T$ (lower panel) operators, for the three DM masses: 100, 150, and 200 GeV as benchmark points.}}
\label{fig:DBN_V}
\end{figure*}


\par Because of their high sensitivity, we focus on the kinematic observables $p_{{T \gamma}}$ and $\eta_\gamma$ when analyzing the form profiles of processes involving mono-photons with missing energy. We produce normalized one-dimensional distributions for the signals created by important operators as well as the background processes of the SM. We depict the one dimensional normalized differential cross sections in Figure \ref{fig:DBN_V} for the three different DM mass values, $100$, $150$, and $200$ GeV, taking into account a center of mass energy of $\sqrt{s}=500$ GeV and an integrated luminosity of 500 $fb^{-1}$ in order to investigate the effect of DM mass.

 The calculation of $\chi^2$ utilizing the two dimensional differential distributions of kinematic observables, $p_{T_\gamma}$ and $\eta_\gamma$, increases the sensitivity of $\Lambda$ to the DM mass. Under the following circumstances, this analysis is carried out for both background and signal processes:

(i) At an integrated luminosity of 250 fb$^{-1}$, for DM mass between 10 GeV and 120 GeV at a center of mass energy, $\sqrt{s} = 250$ GeV.

(ii) Considering an integrated luminosity of 500 fb$^{-1}$, for a DM mass between 10 GeV to 240 GeV at a center of mass energy, $\sqrt{s} = 500$ GeV.
 
The $\chi^2$ is   defined as

\bea{{\chi}}^2  =\sum_{i=1}^{n_1}\sum_{j=1}^{n_2} \left [ \frac{ N_{ij}^{DM}}{\sqrt{  N_{ij}^{SM+DM} +\delta_{\rm sys}^2\ \left({N_{ij}^{SM+DM}}\right)^2}} \right ]^2.
\eea

In this case, the DM and total events inside the two-dimensional grid of ${p_{T_\gamma}}_i$ and ${\eta_\gamma}_j$ are represented by $ N_{ij}^{DM}$ and $ N_{ij}^{SM+DM}$, respectively. The measurement's overall systematic error is denoted by the notation $\delta_{\rm sys}$.

\begin{figure}[h]
\centering
\begin{multicols}{2}
\includegraphics[width=0.49\textwidth]{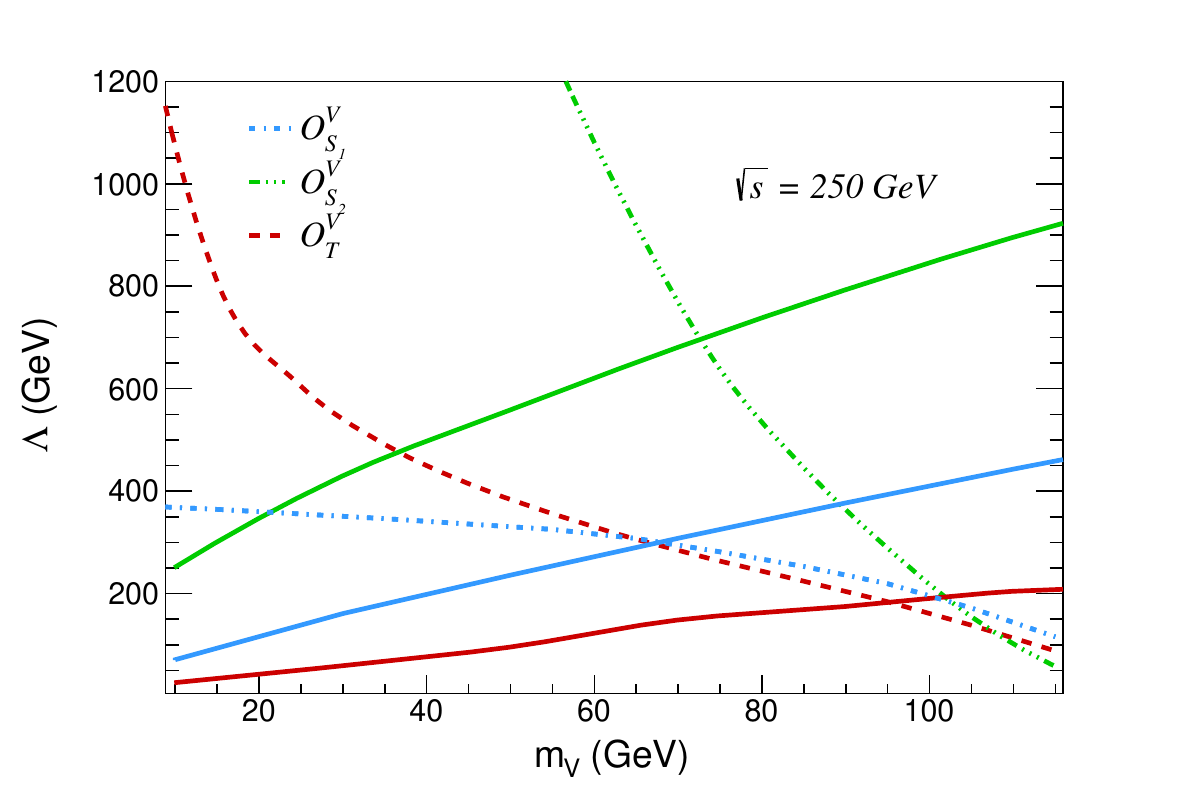}
\columnbreak
\includegraphics[width=0.49\textwidth]{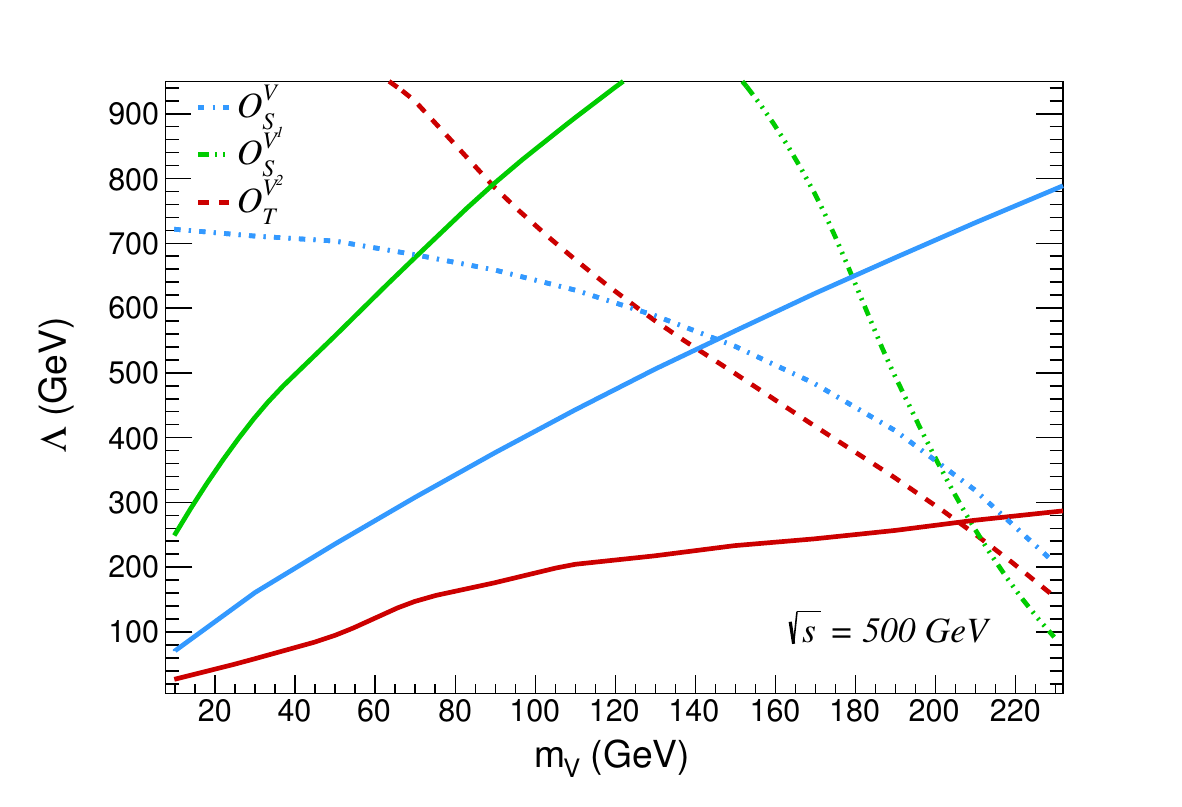}
\end{multicols}
\caption{\small \em {Dashed lines represent the $3\sigma$ contours at a 99.73\% confidence level in the $m_{\text{DM}}-\Lambda$ plane, obtained from $\chi^2$ analyses of the $e^+e^- \to \slash\!\!\! \!E_T +\gamma$ signature at the proposed International Linear Collider (ILC) with center-of-mass energies of $\sqrt{s}$ = 250 GeV and 500 GeV, and integrated luminosities of 250 fb$^{-1}$ and 500 fb$^{-1}$, respectively. The region below each dashed line, corresponding to its respective contour, represents a potentially discoverable area with at least 99.73\% confidence. Additionally, the regions below the solid lines, associated with specific operators, satisfy the relic density constraint $\Omega_{\rm DM}h^2 \le 0.1198$.}}
\label{fig:ILC_V}
\end{figure}   
\unskip

To compute the ${\chi}^2$, we run simulations for two-dimensional differential distributions using the collider parameters listed in Table \ref{table:accelparam} and a cautious estimate of a $1\%$ systematic error. We provide $3\sigma$ contours in the $m_{DM}-\Lambda$ plane at the $99.73\%$ confidence level in Figure \ref{fig:ILC_V}. For effective operators that follow perturbative unitarity, The center of mass energies that these contours represent are $\sqrt{s}=250$ GeV and 500 GeV, respectively..

\section{Summary}
\label{summ}
The phenomenology of DM is examined in this article using an effective field theory as a framework. Neutral electroweak gauge bosons and DM particles are studied in the context of SM gauge-invariant contact interactions up to dimension 8. The study is restricted to self-conjugate DM particles, namely a real vector boson, in order to preserve the invariance of SM gauge symmetry above the electroweak scale. The observed relic density of $\Omega_{DM} h^2=0.1198$ is used to evaluate the relic density contributions of these particles and constrain their characteristics. Furthermore, the operators for the DM mass $> 100$ GeV are severely constrained by comparing the annihilation cross sections for the specified DM mass with indirect detection data from HESS.

The phenomenologically interesting DM mass range of $\le 30$ GeV for $O^V_{S_1}$, $\le 20$ GeV for $O^V_{S2}$, and $\le 22=5$ GeV for $O^V_T$ operators, respectively, is disallowed, according to the analysis, which is extended to existing LEP data. For the pair generation of DM particles at the proposed ILC, a $\chi^2$-analysis is then carried out, encompassing a DM mass between $10-240$ GeV for the effective operators presented in the section \ref{model}. The results show that the dominating mono-photon signal at the future $e^+\ e^-$ collider, ILC, can provide greater sensitivity within the $m_{V}-\Lambda$ parameter space that is constrained by relic density and indirect detection data.

\appendix
\begin{center}
{\bf \Large Appendix}
\end{center}
\section{Thermally averaged cross-sections}
\label{App}

The thermally averaged cross-sectional expressions for the production of photon pair from DM annihilation are computed as follows:
\begin{eqnarray}
\langle \sigma^{V}_{S_1}\ v\rangle\ \left(V\ V \to \gamma \gamma\right)&\simeq&
\frac{16\ {\alpha^{V}_{S_1}}^2\ }{9 \pi}\ \cos^4\theta_w \ \frac{m_{V}^6}{\Lambda^8}\ \left(7\ +\ \frac{10}{x_f}\right)\\
\langle \sigma^{V}_{S_2}\ v\rangle\ \left(V\ V \to \gamma \gamma\right)&\simeq&
\frac{{\alpha^{V}_{S_2}}^2\ }{9\pi}\ \cos^4\theta_w \ \frac{m_{V}^2}{\Lambda^4}\ \left(7\ +\ \frac{10}{x_f}\right)\\
\langle \sigma^{V}_{T}\ v\rangle\ \left(V\ V \to \gamma \gamma\right)&\simeq&
\frac{32\ {\alpha^{V}_{T}}^2\ }{9\pi}\ \cos^4\theta_w \ \frac{m_{V}^6}{\Lambda^8}\ \frac{1}{x_f}
\end{eqnarray}
\par The thermally averaged cross-sectional expressions for the production of $Z$ pair from DM annihilation are computed as follows:
\begin{eqnarray}
\langle \sigma^{V}_{S_1}\ v\rangle\ \left(V\ V \to Z Z\right)&\simeq&
\frac{{\alpha^{V}_{S_1}}^2\ }{9\pi}\ \sin^4\theta_w \ \frac{m_{V}^6}{\Lambda^8}\nn\\
&&\hspace*{-4cm}\times \left[136\ - 128\ \frac{m_V^2}{m_Z^2}\ + 96\ \frac{m_V^4}{m_Z^4}- 80 \frac{m_Z^2}{m_V^2} \ + 24\ \frac{m_Z^4}{m_V^4} +\ \frac{1}{x_f} \left( 384 -752 \frac{m_V^2}{m_Z^2} + 608\ \frac{m_V^4}{m_Z^4} + 16 \frac{m_Z^2}{m_V^2} - 64 \frac{m_Z^4}{m_V^4} \right) \right]\nn \\
\\
\langle \sigma^{V}_{S_2}\ v\rangle\ \left(V\ V \to Z Z\right)&\simeq&
\frac{{\alpha^{V}_{S_2}}^2\ }{9 \pi}\ \sin^4\theta_w \ \frac{m_{V}^2}{\Lambda^4}\nn\\
&&\hspace*{-3cm}\times \left[\frac{17}{2} - 8\ \frac{m_V^2}{m_Z^2}\ + 6\ \frac{m_V^4}{m_Z^4}- 5 \frac{m_Z^2}{m_V^2} \ + \frac{3}{2}\ \frac{m_Z^4}{m_V^4} +\ \frac{1}{x_f} \left( 24 -47 \frac{m_V^2}{m_Z^2} + 38\ \frac{m_V^4}{m_Z^4} +  \frac{m_Z^2}{m_V^2} - 4 \frac{m_Z^4}{m_V^4} \right) \right]\nn \\
\\
\langle \sigma^{V}_{T}\ v\rangle\ \left(V\ V \to Z Z\right)&\simeq&
\frac{{\alpha^{V}_{T}}^2\ }{9 \pi}\ \sin^4\theta_w \ \frac{m_{V}^6}{\Lambda^8}\nn\\
&&\hspace*{-3cm}\times \left[\frac{11}{2} - 2\ \frac{m_V^2}{m_Z^2}\ + 2\ \frac{m_V^4}{m_Z^4}- 4 \frac{m_Z^2}{m_V^2} \ + 3 \frac{m_Z^4}{m_V^4} +\ \frac{1}{x_f} \left( 37 -29 \frac{m_V^2}{m_Z^2} + 26\ \frac{m_V^4}{m_Z^4} - \frac{29}{2} \frac{m_Z^2}{m_V^2} + 3 \frac{m_Z^4}{m_V^4} \right) \right]\nn \\
 \end{eqnarray}
\par The thermally averaged cross-sectional expressions for the production of $Z\ \gamma$ from DM annihilation are computed as follows:
\begin{eqnarray}
\langle \sigma^{V}_{S_1}\ v\rangle\ \left(V\ V \to Z \gamma\right)&\simeq&
\frac{{\alpha^{V}_{S_1}}^2\ }{9 \pi}\ \cos^2\theta_w\ \sin^2\theta_w \ \frac{m_{V}^6}{\Lambda^8}\nn\\
&& \hspace*{-3cm}\times \left[80\ +\ 16\ \frac{m_V^2}{m_Z^2}\ - \frac{63}{2} \frac{m_Z^2}{m_V^2} \ + 2\ \frac{m_Z^4}{m_V^4} +\ \frac{1}{x_f} \left( 123 + 60 \frac{m_V^2}{m_Z^2} - \frac{31}{4} \frac{m_Z^2}{m_V^2} - \frac{13}{4}\ \frac{m_Z^4}{m_V^4} \right) \right]\nn \\
\\
\langle \sigma^{V}_{S_2}\ v\rangle\ \left(V\ V \to Z \gamma\right)&\simeq&
\frac{{\alpha^{V}_{S_2}}^2\ }{9 \pi}\ \cos^2\theta_w\ \sin^2\theta_w \ \frac{m_{V}^2}{\Lambda^4}\nn\\
&& \hspace*{-3cm} \times \left[5\ +\ \frac{m_V^2}{m_Z^2}\ - \frac{63}{32} \frac{m_Z^2}{m_V^2} \ + \frac{1}{8} \frac{m_Z^4}{m_V^4} +\ \frac{1}{x_f} \left( \frac{123}{16} + \frac{15}{4} \frac{m_V^2}{m_Z^2} - \frac{31}{64} \frac{m_Z^2}{m_V^2} - \frac{13}{64}\ \frac{m_Z^4}{m_V^4} \right) \right]\nn \\
\\
\langle \sigma^{V}_{T}\ v\rangle\ \left(V\ V \to Z \gamma\right)&\simeq&
\frac{{\alpha^{V}_{T}}^2\ }{9 \pi}\ \cos^2\theta_w\ \sin^2\theta_w \ \frac{m_{V}^6}{\Lambda^8}\nn\\
&&  \hspace*{-3cm} \times \left[\frac{51}{8}\ + \frac{3}{2} \frac{m_Z^2}{m_V^2} \ + \frac{3}{32} \frac{m_Z^4}{m_V^4} +\ \frac{1}{x_f} \left( \frac{153}{4}\ - \frac{45}{8} \frac{m_Z^2}{m_V^2} + \frac{9}{32}\ \frac{m_Z^4}{m_V^4} \right) \right]
\end{eqnarray}



\end{document}